\documentclass{SCIS2019}

\begin{document}
\Year{2019}
\Month{}
\Vol{}
\No{}
\DOI{}
\ArtNo{}
\ReceiveDate{}
\ReviseDate{}
\AcceptDate{}
\OnlineDate{}

\title{Photonic Integrated Phase Decoder Scheme for High-Speed, Efficient and Stable Quantum Key Distribution System}{Title for citation}

\author[1]{Huaxing XU}{{andy{\_}huaxingxu@126.com}}
\author[1]{Yaqi SONG}{}
\author[2]{Xiaofan MO}{}
\author[3]{Yitang DAI}{}
\author[1]{\\Changlei WANG}{}
\author[1]{Shaohua WANG}{}
\author[1]{Rui ZHANG}{}

\AuthorMark{Xu H X}



\address[1]{National Engineering Laboratory for Public Safety Risk Perception and Control by Big Data, \\China Academy of Electronics and Information Technology, Beijing {\rm 100041}, China}
\address[2]{National Astronomical Observatories, Chinese Academy of Sciences, Beijing {\rm 100012}, China}
\address[3]{State Key Laboratory of Information Photonics and Optical Communications, \\ Beijing University of Posts and Telecommunications, Beijing {\rm 100876}, China}

\abstract{Quantum key distribution (QKD) is gradually moving towards network applications. It is important to improve the performance of QKD systems such as photonic integration for compact systems, the stability resistant to environmental disturbances, high key rate, and high efficiency in QKD applications. In the letter, we propose a general quantum decoding model, namely orthogonal-polarizations-exchange reflector Michelson interferometer model, to solve quantum channel disturbance caused by environment. Based on the model, we give a quantum phase decoder scheme, \emph{i.e.} a Sagnac configuration based orthogonal-polarizations-exchange reflector Michelson interferometer (SRMI). Besides the stability immune to quantum channel disturbance, the SRMI decoder can be fabricated with photonic integrated circuits, and suitable to gigahertz phase encoding QKD systems, and can increase the system efficiency because of the low insertion loss of the decoder.}


\maketitle

\section{Introduction}
Quantum key distribution (QKD) \cite{gisin2002quantum} allows two authenticated distant participants Alice and Bob to share a long random string often called cryptographic keys with information theoretic security. The keys can be used to carry out perfectly secure communication via one-time-pad.
The first QKD protocol was proposed by Bennett and Brassard in 1984 \cite{BB84}. Then significant progresses on QKD schemes \cite{E91,6state,SARG04} and security proofs have been made. QKD has been strictly proved to be unconditionally secure and composably secure \cite{Mayers,Lo,Shor,Renner,Ben2005The}, which means that one-time-pad algorithm using the keys distributed by QKD can resist eavesdroppers with unbounded abilities. Practical security of QKD has also been fully studied, such as the decoy state method \cite{hwang2003quantum,wang2005decoy,lo2005decoy} for beating the photon-number-splitting (PNS) attack aiming at weak coherent source, which now has been applied in common QKD systems, and the measurement-device-independent (MDI) QKD for removing detector side channel attacks \cite{lo2012measurement}.

With the developments over the past three decades, the demonstration of QKD has been processed with longer distance. However, the instability of QKD systems caused by the optical fiber quantum channel and low key generation rate limit the applications of QKD. In terms of the encoding type, QKD contains the polarization encoding systems, phase encoding systems, and time-bin phase encoding systems, where time-bin phase encoding includes a set of phase basis and a set of time-bin basis. The polarization encoding QKD systems via optical fiber can be easily realized in a laboratory but relies on complicated feedback compensation, which is not suitable for the quantum channel with strong environmental disturbance.
Since the phase information can be maintained in environmental disturbance, phase encoding or time-bin phase encoding QKD systems are more competitive over overhead and tube optical cable along roads or bridges. The phase encoding systems are mainly based on asymmetric Mach-Zehnder interferometers (AMZI) \cite{B92}, which suffer from polarization induced fading in transmission fiber and result in the fringe visibility of the interferometers varying fast.
To solve the problem, A. Muller \emph{et al.} proposed the ``plug and play'' bidirectional QKD system which can automatically compensate the polarization disturbance besides the phase drifting in the channel \cite{muller1997plug}, and has been applied in QKD products by ID Quantique, Inc., a well-known Swiss quantum company. But trojan-horse attack on ``plug and play'' systems was proposed by N. Gisin  \emph{et al.} in 2006 \cite{gisin2006trojan}.
Then X. F. Mo \emph{et al.} proposed the Faraday-Michelson (F-M) unidirectional QKD system \cite{Xiao2005Faraday}, which has been applied in the phase encoding QKD products by Anhui Asky Quantum Technology Co., Ltd., and designed in the latest timing-bin phase QKD scheme in the patent proposed by University of Science and Technology of China \cite{Patents}. On the premise of maintaining QKD system against environment mutation, a gigahertz QKD scheme based on Faraday-Michelson-Sagnac interferometer (FMSI) was proposed in Ref. \cite{wang2018practical}, which achieves the efficient and high key generation rates by decreasing the insertion losses and increasing the phase modulation rates. In FMSI scheme, a Sagnac configuration with a $90^\circ$ Faraday rotator and a phase modulator (PM) replaces the Faraday mirror in one arm of the Michelson interferometer.
Up to now, the effective and widely used solutions are mainly based on ``plug and play'' and F-M schemes or their variants. Both the two solutions are based on the same principle in essence because they both take advantage of the Faraday magneto-optic effect. In the future, miniaturization is the developing trend of QKD technology. However, the schemes with the Faraday magneto-optic effect are difficult to be designed by the integrated optical waveguide.

In this paper, we propose an orthogonal-polarizations-exchange reflector Michelson interferometer (OPERMI) model, which does not need a Faraday rotator. All the QKD decoder satisfying OPERMI model can be free of polarization disturbances caused by the quantum channel. Since for phase encoding QKD systems the disturbances of quantum channel will be collected in the system if there is polarization-induced fading at the receiver¡¯s interferometer\cite{han2005stability}, the model of interferometer can be self-compensating quantum channel disturbance. Then we put forward a Saganc configuration based orthogonal-polarization-exchange reflector Michelson interferometer (SRMI) as the decoder without Faraday elements based on OPERMI model, which is a photonic integrated phase decoder scheme for high-speed, efficient and stable QKD system.
In our decoder scheme, it realizes the exchange of the orthogonal polarizations by a PBS based Sagnac configuration. Setting the PM in the middle of the Sagnac loop, the insertion losses of the decoder decreases and the PM can be modulated to gigahertz speed. Our scheme is simple in realization and all the components in the scheme are commercial passive optical components, which can be easily fabricated. Besides, our decoder is easy to be integrated, especially the Sagnac configuration reflector. The photonic integrated phase decoder scheme is designed in Section 4, which can be applied in the miniaturized QKD industry in the future.

\section{Model of Orthogonal-Polarizations-Exchange Reflector Michelson Interferometer}
In this section, we extract a general model based on Michelson interferometer from our previous work in Ref. \cite{Xu2019} (denoted as quarter-wave plate reflector Michelson interferometer scheme, \emph{i.e. }Q-M scheme) to realize the stable phase encoding QKD system.

\begin{figure}[!t]
\centering
\begin{minipage}[c]{0.48\textwidth}
\centering
\includegraphics{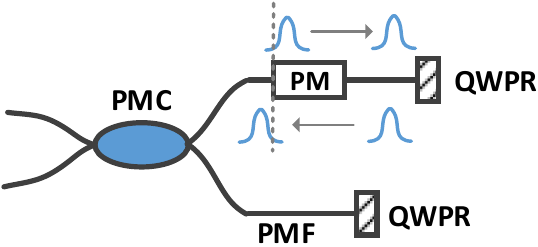}
\end{minipage}
\hspace{0.02\textwidth}
\begin{minipage}[c]{0.48\textwidth}
\centering
\includegraphics{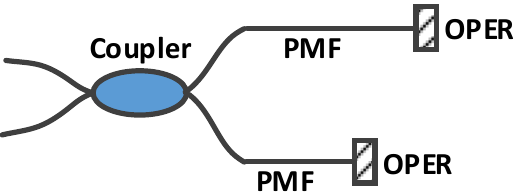}
\end{minipage}\\[3mm]
\begin{minipage}[t]{0.48\textwidth}
\centering
\caption{Schematic of QMI. The asymmetric Michelson interferometer contains a PMC, two unbalanced arms, a PM in the upper arm and two QWPRs.}
\label{fig:QM scheme}
\end{minipage}
\hspace{0.02\textwidth}
\begin{minipage}[t]{0.48\textwidth}
\centering
\caption{Model of OPERMI. It contains a coupler, two unbalanced arms and two OPERs}
\label{fig:OPERMI}
\end{minipage}
\end{figure}

According to Ref. \cite{Xu2019}, the Q-M interferometer (QMI) is composed of a polarization-maintaining coupler (PMC), two unbalanced arms consisted of polarization-maintaining fiber (PMF), and two quarter-wave plate reflectors (QWPRs), shown in \figurename~\ref{fig:QM scheme}. Besides, there is a phase modulator (PM) in one of the two arms, in the upper or the lower arm. The arbitrary polarization input light in the QMI can be treated as two orthogonal polarizations parts transferred along the slow and fast axis of the PMF arms, respectively. After reflected by QWPR, the forward part along the slow axis of the PMF arms is turned to the backward light along the fast axis of the PMF arms, and vice versa. Due to the same phase accumulation during the round-trip transmission, only two orthogonal polarization state exchange happens between the input and output light, namely, the output polarization state is independent of the two arms. Therefore, QMI can eliminate the polarization-induced signal fading caused by the interferometer and quantum channel disturbance can also be automatically compensated.
%
Q-M scheme can be seen as an implementation of a general model, which we call it the orthogonal-polarizations-exchange reflector Michelson interferometer (OPERMI). All QKD systems with OPERMI can automatically compensate the channel polarization disturbance. As shown in \figurename~\ref{fig:OPERMI}, it is composed of a coupler, two unbalanced arms and two reflectors. The model of OPERMI has the following two key points.
\begin{itemize}
    \item The two arms of the interferometer maintain the two orthogonal polarizations unchanged during transmission.
    \item The reflector exchanges the two orthogonal polarizations when reflecting them back. We call the reflectors as orthogonal-polarizations-exchange reflectors (OPERs).
\end{itemize}

A simple proof is given to explain why the OPERMI can be self-compensating quantum channel disturbance. The function of the OPERs is to exchange the two orthogonal polarizations and reflects them back. That is, a state $|\Psi\rangle=a|H\rangle+b|V\rangle$ is changed to be $|\Psi'\rangle=b|H\rangle+a|V\rangle$ by OPERs. Consider that the two arms of the interferometer maintain the two orthogonal polarization unchanged. Therefore, the operator $U$ of OPERMI is
\begin{equation}
\begin{aligned}
    \begin{bmatrix}
      b\\a
    \end{bmatrix}
   &=
    U
    \begin{bmatrix}
      a\\b
    \end{bmatrix}
   \\
    U&=
     \begin{bmatrix}
      0 & 1 \\ 1 & 0
    \end{bmatrix}
    =\sigma_2.
\end{aligned}
\end{equation}
It can be seen that $U^\dag U=I$, which satisfies the stable condition of phase-modulated QKD system according to Ref. \cite{han2005stability}. To explain the stability in detail, the transformation operations are given as follows. Take the arms of the OPERMI as PMFs for example, which maintain the two orthogonal polarizations unchanged during transmission. A beam of light passes though the long arm of the OPERMI and is reflected by a OPER. The operator $L$ of the long arm is
\begin{equation}
  \begin{aligned}
    L&=\stackrel{\leftarrow}{l}\cdot U\cdot \stackrel{\rightarrow}{l}\\
    &= \begin{bmatrix} exp(i\delta/2) & 0 \\ 0 & exp(-i\delta/2)    \end{bmatrix}
         \begin{bmatrix}  0 & 1 \\ 1 & 0    \end{bmatrix}
         \begin{bmatrix} exp(i\delta/2) & 0 \\ 0 & exp(-i\delta/2)    \end{bmatrix}\\
    &= \begin{bmatrix}  0 & 1 \\ 1 & 0    \end{bmatrix}= \sigma_2,
  \end{aligned}
\end{equation}
where $l$ is the operator of PMF in the long arm, $\leftarrow$ and $\rightarrow$ indicated the backward and forward propagation, $\delta$ is the birefringence strength of the PMF. The conclusion also applies to the short-arm operator $S$.

\begin{figure}[htbp]
\centering
\includegraphics[width=0.8\textwidth]{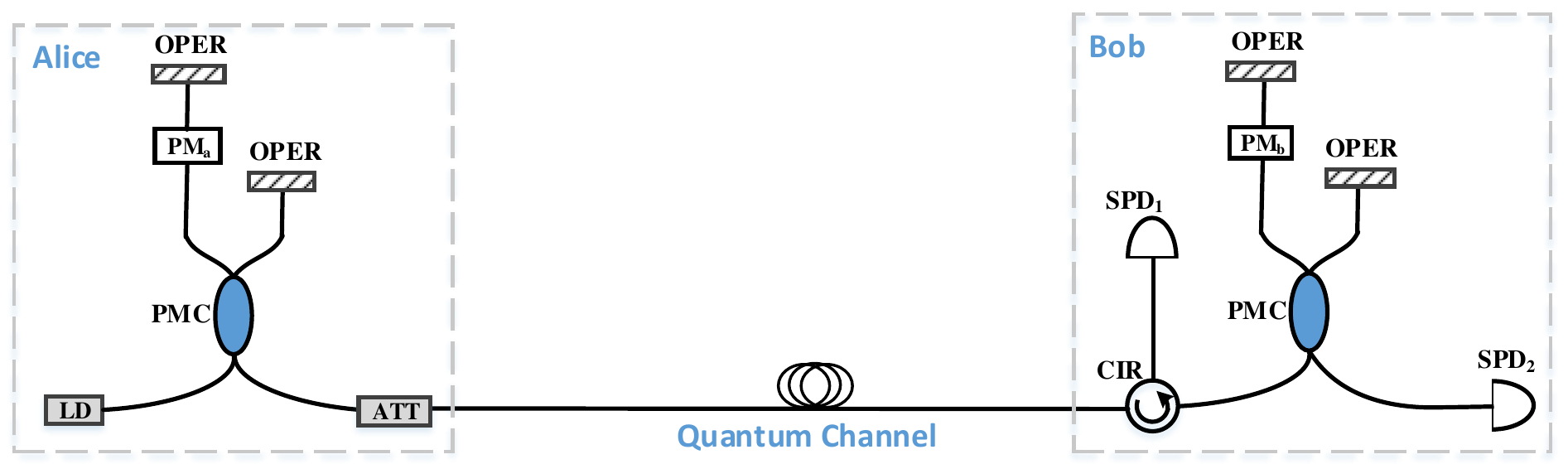}
\caption{Schematic of phase encoding QKD system with two OPERMIs. LD is a laser dopde. ATT is an attenuator. CIR is an optical circulator. SPDs are two single-photon detectors. $PM_a$ and $PM_b$ are the phase modulators at Alice's and Bob's side, respectively.}
\label{fig:wholesystem}       
\end{figure}

Then analyze the operator of the whole system with two OPERMIs at Alice and Bob as the quantum encoder and decoder, respectively. The schematic of the whole system is shown in \figurename~\ref{fig:wholesystem}. There are two paths leading the single-photon interference. One path contains the long arm of Alice's OPERMI, the quantum channel and the short arm of Bob's OPERMI while the other contains the short arm on Alice's side, the quantum channel and the long arm on Bob's side. Suppose two PM are both on the long arm of OPERMI. The paths are described as
\begin{equation}
  \begin{aligned}
    &Path 1: L_a\rightarrow PM_a\rightarrow QC\rightarrow S_b,\\
    &Path 2: S_a\rightarrow QC\rightarrow L_b\rightarrow PM_b,
  \end{aligned}
\end{equation}
where QC represents the Jones matrix of quantum channel and the subscripts $a$ and $b$ represents the operators of the components at Alice's and Bob's side, respectively. Then the transformation matrices of two paths are
\begin{equation}
  \begin{aligned}
    T_1&=\left[exp(i\varphi_{s_b})\cdot\stackrel{\longleftarrow}{s_b}\cdot U\cdot \stackrel{\longrightarrow}{s_b}\right]\cdot \left[exp(i\varphi_{c})\cdot QC\right]\cdot \left[exp(i\varphi_{l_a}+i\varphi_{a})\cdot \stackrel{\longleftarrow}{PM_a}\cdot\stackrel{\longleftarrow}{l_a}\cdot U\cdot \stackrel{\longrightarrow}{l_a}\cdot \stackrel{\longrightarrow}{PM_a}\right]\\
    &=exp[i(\varphi_{s_b}+\varphi_{c}+\varphi_{l_a}+\varphi_{a})]\cdot \sigma_2\cdot QC\cdot \sigma_2,\\
    T_2&= \left[exp(i\varphi_{l_b}+i\varphi_{b})\cdot \stackrel{\longleftarrow}{PM_b}\cdot\stackrel{\longleftarrow}{l_b}\cdot U\cdot \stackrel{\longrightarrow}{l_b}\cdot \stackrel{\longrightarrow}{PM_b}\right]\cdot \left[exp(i\varphi_{c})\cdot QC\right]\cdot \left[exp(i\varphi_{s_a})\cdot\stackrel{\longleftarrow}{s_a}\cdot U\cdot \stackrel{\longrightarrow}{s_a}\right]\\
    &=exp[i(\varphi_{s_a}+\varphi_{c}+\varphi_{l_b}+\varphi_{b})]\cdot \sigma_2\cdot QC\cdot \sigma_2,
  \end{aligned}
\end{equation}
where $\varphi_{s_a}$, $\varphi_{s_b}$, $\varphi_{l_a}$, $\varphi_{s_l}$ and $\varphi_{c}$ are propagation phases of corresponding components, $\varphi_{a}$ and $\varphi_{b}$ are modulated phases operated by Alice and Bob, respectively. Supposing the input Jones vector of OPERMI on Alice's side is $E_{in}$, the output of Bob's interferometer is
\begin{equation}
  E_{out}=\frac{1}{4}\{exp[i(\varphi_{s_b}+\varphi_{c}+\varphi_{l_a}+\varphi_{a})] +exp[i(\varphi_{s_a}+\varphi_{c}+\varphi_{l_b}+\varphi_{b})]\}\cdot \sigma_2\cdot QC\cdot \sigma_2\cdot E_{in},
\end{equation}
Considering that $C$ is unitary, the interference output power can be expressed as
\begin{equation}
  \begin{aligned}
    P_{out}&=E_{out}^\dag\cdot E_{out}\\
    &=\frac{1}{8}[1+\cos(\Delta\varphi_a+\Delta\varphi_b+\Delta\varphi_{ab}))]P_{in},
  \end{aligned}
\end{equation}
where $\Delta\varphi_a=\varphi_{l_a}-\varphi_{s_a}$, $\Delta\varphi_b=\varphi_{l_b}-\varphi_{s_b}$, and $\Delta\varphi_{ab}=\varphi_{a}-\varphi_{b}$. It means that the interference output $P_{out}$ is independent of any polarization perturbation in the whole QKD system, especially that caused by quantum channel. That is, the OPERMI model can make QKD system free of polarization disturbances caused by the quantum channel.


\section{Sagnac Configuration Based Orthogonal-Polarizations-Exchange Reflector Michelson Interferometer Scheme}
There exists a common problem of the asymmetric Michelson-interferometer phase encoding scheme to support high-speed QKD systems \cite{2012Exp}. Taking Ref. \cite{Xu2019} as an example, the light path of the long arm is $PM\rightarrow QWPR\rightarrow PM$, as seen in \figurename~\ref{fig:QM scheme}. The settling time of PM is during the forward input to the PM and the backward output from the PM of a pulse. The modulating voltage should remain constant during the state transferred through the PM twice.
The settling time of PM limits the speed of the whole system. For a fixed phase modulation time, the longer settling time determines a shorter time of jump edge. To achieve a shorter jump edge time, ultra-high-speed digital-to-analog converter (DAC) is needed. Therefore, the asymmetric Michelson interferometer based high-speed QKD systems have a high requirement for DAC. Besides, the insertion loss of the PM (about 4dB) is the main losses of the interferometer. Due to the modulated pulse passing through the PM twice, double insertion loss is introduced by the PM, resulting low efficiency of the system.

\begin{figure}[htbp]
\centering
\includegraphics[width=0.5\textwidth]{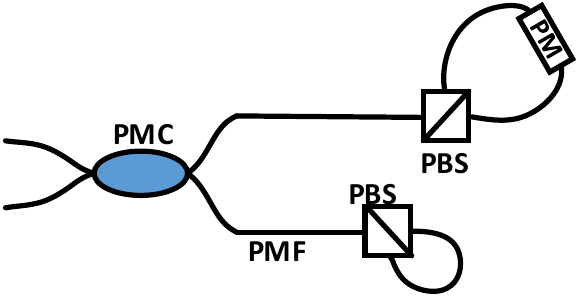}
\caption{Schematic of the complete SRMI.}
\label{fig:CSRMI}       
\end{figure}

We propose a high-speed, efficient phase decoder scheme of QKD based on the OPERMI model, which we call it Sagnac configuration based orthogonal-polarizations-exchange reflector Michelson interferometer (SRMI). As shown in \figurename~\ref{fig:CSRMI}, the SRMI is composed of a PMC, two PMF arms and two Sagnac configuration reflectors as OPERs. The Sagnac configuration reflector is made up by a PBS and a $90^\circ$ twisted  PMF, where the two output ports of PBS are connected with a $90^\circ$ twisted PMF, and both two ports can be coupled to the slow (or fast) axis of the $90^\circ$ twisted PMF. In the scheme, PM can be joined up with the PMF in the Sagnac configuration reflector for phase encoding. For more convenient application, the PM is located in the middle of the Sagnac configuration. When a pulse is transferred trough the PBS and divided into two orthogonal parts, the two parts are modulated by the PM simultaneously.

\begin{figure*}[htbp]
\centering
\includegraphics[width=0.9\linewidth]{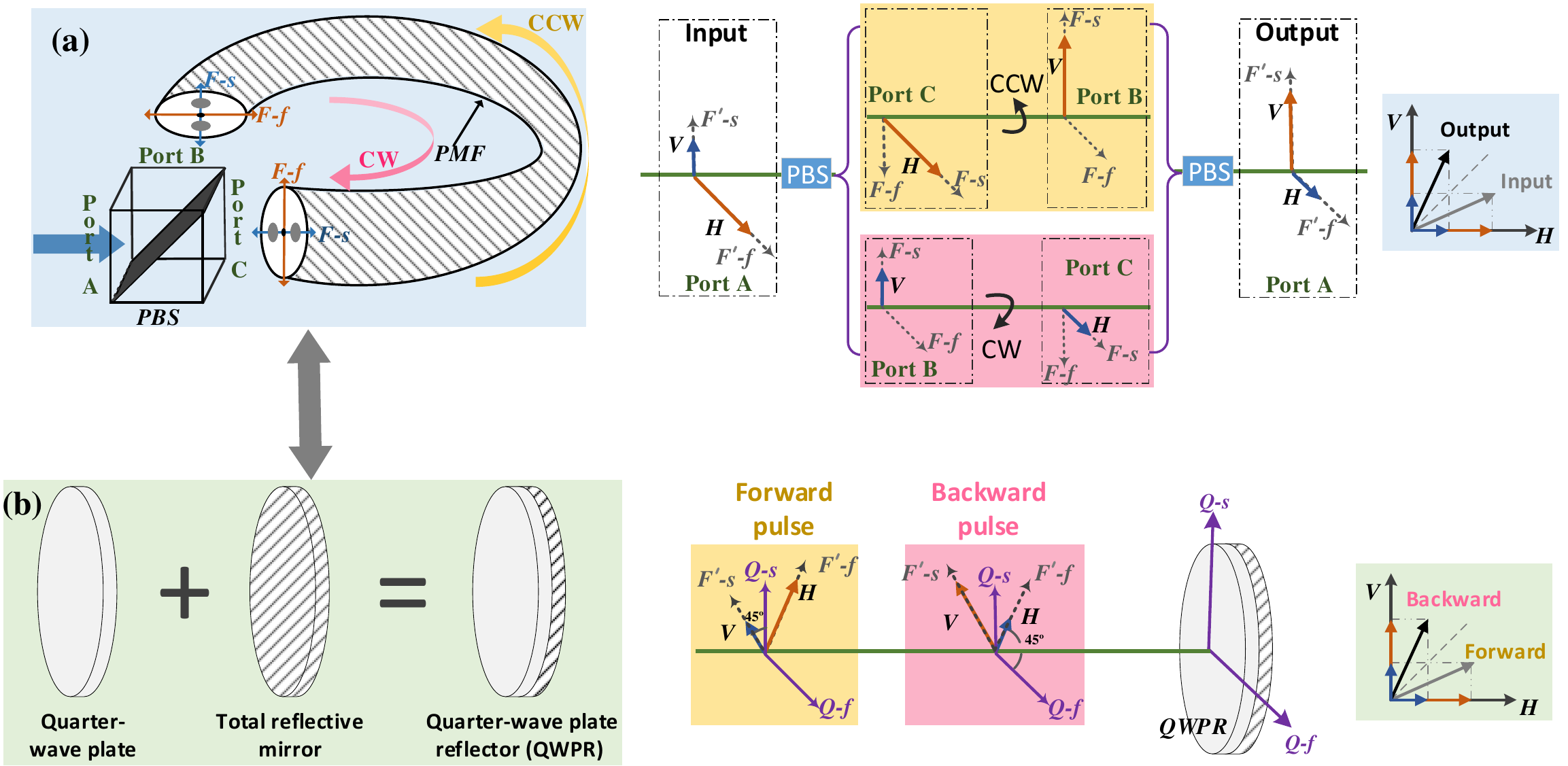}
\caption{The equivalence of the Sagnac configuration and QWFR. Both two structures achieve the exchange of $H$ and $V$ in the situation that the phase accumulations of two polarizations are the same. $F'$-$s$ and $F'$-$f$ are the slow and fast axes of the PMF belong to the long arm of Mechelson interferometer, respectively. $F$-$s$ and $F$-$f$ are the slow and fast axes of the PMF belong to Sagnac loop, respectively. $Q$-$s$ and $Q$-$f$ are the slow and fast axes of QWP, respectively. The angle between the slow axis of PMF and QWP is $45^\circ$.}
\label{fig:S=Q}       
\end{figure*}

The Sagnac configuration reflector exchange the two orthogonal polarizations along the slow (or fast) axis of the $90^\circ$ twisted PMF. The Sagnac configuration reflector has the same function with QWPR, where the horizontal polarization is turned to the vertical polarization and the vertical polarization is turned to the horizontal polarization. They both achieve the exchange of the orthogonal polarizations. The configuration and the process of the two components are shown in \figurename~\ref{fig:S=Q}. In the Sagnac configuration, the input port of PBS is denoted as $Port$ $A$ while the output ports are denoted as $Port$ $B$ and $Port$ $C$, respectively. Here assume both $Port$ $B$ and $Port$ $C$ are coupled to the slow axis of the twisted PMF (denoted as $F$-$s$ axis). When an arbitrarily polarized pulse passes the PBS, it is divided into two parts. One part is the horizontally polarized pulse ($H$), where the polarization is along $F$-$s$ axis, output at $Port$ $C$ and transferred along the counter clockwise (CCW) direction. The other part is the vertically polarized pulse ($V$), where the polarization is along $F$-$s$ axis, output at $Port$ $B$ and transferred along the clockwise (CW) direction. Taking the horizontal polarization as an example, the direction of $H$ is in line with $F$-$s$ axis at $Port$ $C$ because the port is coupled to the slow axis of PMF. To make $Port$ $B$ also coupled with the slow axis of PMF, the PMF must be twisted. With the twist of the PMF and the rotation of coordinate system, the polarization of the pulse is gradually changed. Finally, the pulse $H$ is converted to $V$ at $Port$ $B$. Similarly along the CW direction, the pulse $V$ at $Port$ $B$ is turned to $H$ at $Port$ $C$ by the twisted PMF. Due to the same phase accumulation of the CW-direction path and the CCW-direction path, only polarization exchange happened in the Sagnac configuration. In \figurename~\ref{fig:S=Q}(b), QWPR is composed of a quanter-wave plate (QWP) and a total reflective mirror. The angle between the slow axes of PMF (denoted as $F'$-$s$ axis) and QWP (denoted as $Q$-$s$ axis) is $45^\circ$. A forward $V$ ($H$) polarization pulse along the $F'$-$s$ ($F'$-$f$) axis can be transformed into a backward $H$ ($V$) polarization output light along the $F'$-$f$ ($F'$-$s$) axis after the reflection by QWPR. Thus, the Sagnac configuration reflector has the same function with QWFR.

\begin{figure}[h]
\centering
\includegraphics[width=0.5\textwidth]{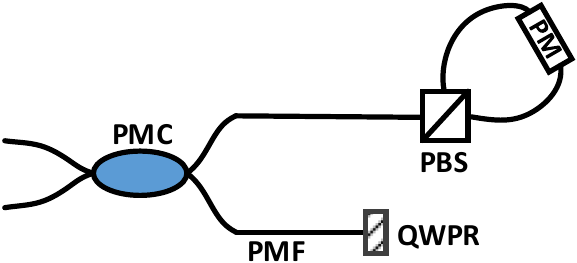}
\caption{Schematic of the simplified SRMI. }
\label{fig:SSRMI}       
\end{figure}

The Jones matrix of both Sagnac configuration reflector and QWPR is $\sigma_2$, which satisfies the theory of OPERMI. The structure in \figurename~\ref{fig:CSRMI} can be simplified as the structure in \figurename~ \ref{fig:SSRMI}, because the equivalence of the Sagnac configuration reflector and QWPR. And it can automatically eliminate the polarization-induced signal fading.

In \figurename~\ref{fig:QM scheme}, the PM needs a birefringent PM. For the birefringent PM, the modulated phase in the horizontal direction is different from that in the vertical direction.  When a pulse first passes through the PM, a $\phi_0$ phase is added to the horizontal direction while a $\phi_1$ phase is added to the vertical direction, where $\phi_0\neq \phi_1$. After reflected by QWPR, the orthogonal polarizations are exchanged and they passes through the PM for the second time. The original horizontal polarization is turned to the vertical polarization and modulated a $\phi_1$ phase, while the original vertical polarization is turned to the horizontal polarization and modulated a $\phi_0$ phase.  Then the unitary pulse is added a phase $\phi=\phi_0+\phi_1$.
In the SRMI scheme, which is shown in \figurename~\ref{fig:SSRMI}, the PM is located in the middle of Sagnac loop. When a pulse goes into the modulation arm, it is divided into two parts by PBS, \emph{i.e.} $H$ transmitting CCW along the $90^\circ$ twisted PMF from $Port C$ to $Port B$ while $V$ transmitting CW along the $90^\circ$ twisted PMF from $Port B$ to $Port C$. Both output ports of the PBS are coupled to the slow (or fast) axis of the $90^\circ$ twisted PMF, and the orthogonal polarization parts are modulated with the same phase $\phi$ by the PM. Then SRMI scheme has the same phase modulation effects with Q-M scheme. Both birefringent PM and single-polarization PM can be used in SRMI scheme. In addition, the modulation voltage in SRMI scheme only needs to stay constant only during a pulse width, since the states from two directions arrive at the PM simultaneously. The general DAC technology can provide the stable modulation voltage for the gigahertz SRMI scheme. Due to the modulated pulse passing through the PM only once, the insertion loss is decreased, which can realize a high efficient QKD decoder.

\section{Integrated Waveguide Scheme for SRMI}

\begin{figure}[!h]
\centering
\includegraphics[width=0.5\textwidth]{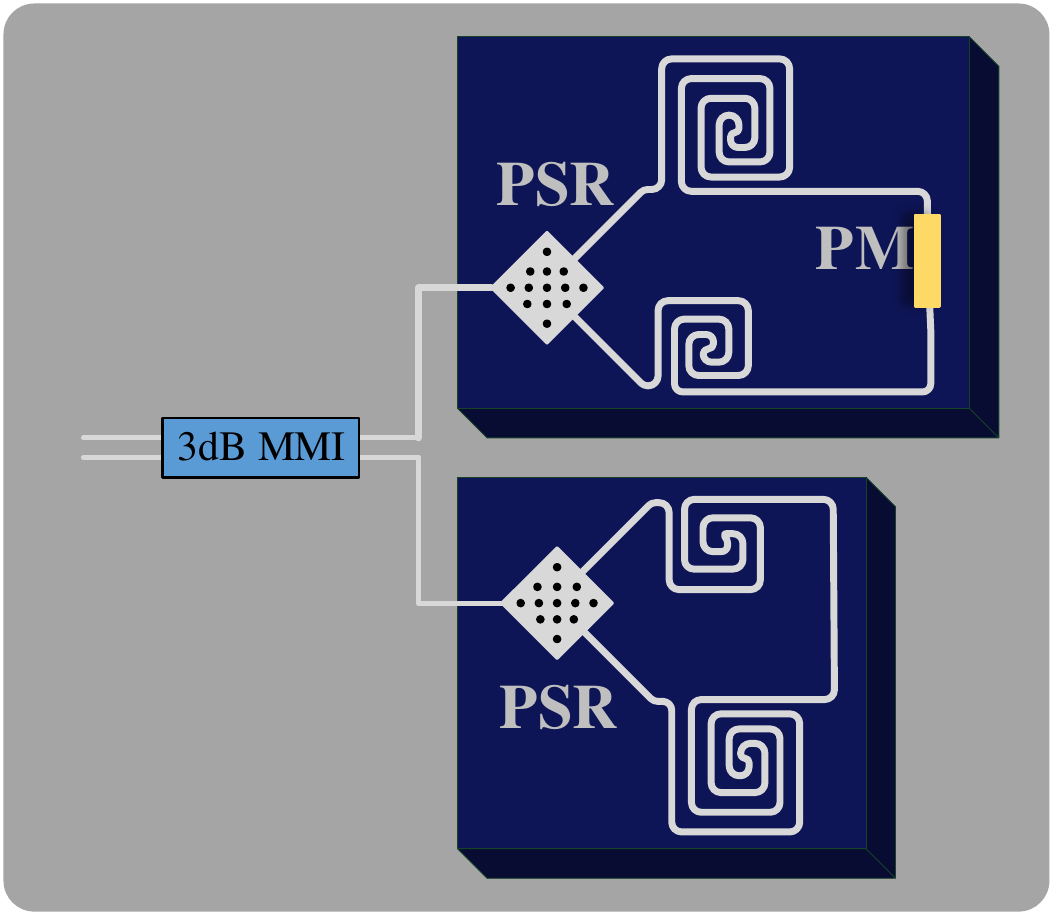}
\caption{Integrated Sagnac configuration reflector.}
\label{fig:Intergration}       
\end{figure}

Each component used in SRMI is easy to be designed by using integrated optical waveguide, especially the Sagnac configuration reflector. The integrated waveguide scheme for SRMI is shown in \figurename ~\ref{fig:Intergration}. The integrated Sagnac configuration reflector is composed of the polarization splitter and rotator (PSR), optical waveguide delay and PM. PSR separates the input light to the  transverse-electric (TE) and transverse-magnetic (TM) modes along two different paths, \emph{i.e.} CW and CCW directions, and transform the TM mode to TE mode. Then the TE mode in two paths can transmit along the optical waveguide of the Sagnac loop. The optical waveguide delay can be designed in the Saganc loop. The beam splitter is realized by a 3dB multi-mode interferometer (MMI), which supports the transformation for both TE and TM modes. The length of the waveguide between the 3dB MMI and PSR is  short enough to make both TE and TM modes transmit. Then the encoder/decoder based on the SRMI can be miniaturized by using integrated optical waveguide.

\section{Conclusions}

We put forward an OPERMI model as the phase decoder for QKD system to be free of polarization induced fading caused by quantum channel. Then we propose a photonic integrated SRMI phase decoder scheme based on OPERMI model. Compered with the QKD scheme in Ref. \cite{Xiao2005Faraday,Xu2019}, SRMI scheme has an advantage in the high efficiency and high-speed modulation while keeping the stability. Compared with another high-speed QKD scheme, seen in Ref. \cite{wang2018practical}, where Faraday rotator is one of the key components and is difficult to be designed by integrated optical waveguide, our scheme is superior in photonic integration for gigahertz QKD systems with high efficiency. This work is significant for the practical applications and miniaturization of QKD industry.

\Acknowledgements{This work was supported by National Natural Science Foundation of China (Grant Nos. 61705202) and the Innovation Funds of China Academy of Electronics and Information Technology.}





\end{document}